\documentclass[10pt, conference, compsocconf]{IEEEtran}
\usepackage{graphicx}
\usepackage{amsmath}
\usepackage{subfigure}
\usepackage{rotating}
\usepackage{multirow}
\usepackage{array}
\usepackage{algorithm}
\usepackage{algorithmic}

\begin{document}

\title{ MODLEACH: A Variant of LEACH for WSNs}

\author{D. Mahmood$^{1}$, N. Javaid$^{1}$, S. Mahmood$^{2}$, S. Qureshi$^{3}$, A. M. Memon$^{4}$, T. Zaman$^{5}$\\
        $^{1}$COMSATS Institute of Information Technology, Islamabad, Pakistan.\\
        $^{2}$Netsolace,Islamabad, Pakistan.\\
        $^{3}$Hamdard University, Islamabad, Pakistan.\\
        $^{4,5}$IT Section, APP, $^{4}$Islamabad,$^{5}$Karachi, Pakistan.\\
     }

\maketitle
\begin{abstract}

Wireless sensor networks are appearing as an emerging need for mankind. Though, Such networks are still in research phase however, they have high potential to be applied in almost every field of life. Lots of research is done and a lot more is awaiting to be standardized. In this work, cluster based routing in wireless sensor networks is studied precisely. Further, we modify one of the most prominent wireless sensor network's routing protocol ``LEACH'' as modified LEACH (MODLEACH) by introducing \emph{efficient cluster head replacement scheme} and \emph{dual transmitting power levels}. Our modified LEACH, in comparison with LEACH out performs it using metrics of cluster head formation, through put and network life. Afterwards, hard and soft thresholds are implemented on modified LEACH (MODLEACH) that boast the performance even more. Finally a brief performance analysis of LEACH, Modified LEACH (MODLEACH), MODLEACH with hard threshold (MODLEACHHT) and MODLEACH with soft threshold (MODLEACHST) is undertaken considering metrics of throughput, network life and cluster head replacements.
\end{abstract}

\begin{IEEEkeywords}
LEACH, Wireless, Sensor, Networks, Routing, Protocol, MODLEACH, WSN's, Cluster, Head, Threshold
\end{IEEEkeywords}

\section{Introduction}

Information- information- information; we need immediate information in every aspect of our lives. A mechanical engineer wants to get information about his machine running, an electrical engineer needs to know about his operating circuitry, gradual changes in earth and climate are required by geologists; list goes on. To full fill this need, several networks are designed to pass information. Initially man to man, than telephonic systems, further giving birth to cellular systems. Computer networks run parallel. Ad-Hoc networks give infrastructure-less communication. Multi hop networks were designed to give more liberty of movement. And then comes the era where machine to machine communication was initiated. A processing device acquires information, process it and transfer it to another machine or processing device. That information is further precised/ aggregated/ fused intelligently so that it can be presented to us ``humans''. In wireless sensor networks, that device normally is termed as a sensor, node or mote and has its own limitations i.e. it must be capable of sensing, processing and transmitting/ receiving. Each node hence also require a power source to do all these operations. Considering applications of wireless sensor networks, installing a battery on each sensor node is a better option. However, limiting use of power is one of the key challenges in wireless sensor networks. These batteries must be smart enough to give a node maximum life despite of being tiny sized. \\
Any technology that is in process of its development, give a lot of challenges. In the same way, wireless sensor networks do. Sensing, computing and transcieving by tiny sized sensors with power constraint is not a simple thing. Hence this is the major concern for scientists and researchers. To optimize node's life time, we need to focus on such algorithms, protocols and physical circuitries that can make maximum out of limited power source.\\
In any network especially wireless multi hop networks, for efficient performance, its protocols must be very efficient. Numerous protocols are developed that address power problem in sensor networks. Most prominent routing algorithms can be categorized into three types i.e direct transmission algorithms, hop to hop transmission algorithms and cluster based algorithms.\\
Another problem that persists is to handle bulk of information sensed and passed over by every node of a network. (A WSN may consist of thousands of nodes). For that data aggregation and data fusion algorithms work, however there is always a room for betterment. In an efficient wireless sensor network, we need efficient routing protocol that has low routing overhead and well organized data aggregation mechanisms to increase good put of network and to save limited power of sensor node. \\
In next sections, we discuss about the work done on cluster based routing of wireless sensor networks along with areas which need modifications to enhance efficiency. Later, some modifications are made in one of most prominent routing protocol. Finally, experiments along with comparisons are made and discussed briefly.

\section{Related Work}
Manufacturing of cheap wireless sensor nodes having sufficient computation and transmitting/ receiving powers are available now. Hence hundreds of nodes can be deployed in a network for any required application. These sensor nodes have a limited power which must be utilized in very precise manner to increase node's life. No doubt efficient circuit is necessary for efficient use of energy, however, routing protocol running on the network plays a vital role in bandwidth consumption, security and energy conservations as well (considering WSN's).\\
To cop with these constraints, initially direct transmission approach was discussed [1]. In direct transmission, a node sense data from its environment and transmits it straight to base station. This method, no doubt, ensures data security however, on the other hand we have to compromise on node's life time due to excessive power consumption (if BS is far away). Hence, using direct transmission technique, nodes that are far away from BS die early as they require more power to propagate their signal, making a portion of field vacant for sensing.\\
To solve this problem, minimum transmission energy (MTE) emerged. In this technique, data is transmitted to base stations via multi hop. This gives birth to almost same problem we faced in direct transmission. Difference is only this that in minimum transmission energy algorithm, far away nodes remain alive longer with respect to the nodes nearer to BS. Reason behind early expiry of nearer nodes is routing of all data traffic to base station. More over, transmitting bulk of sensed data from each node use much energy. To overcome this problem, concept of Directed Diffusion was introduced that discuss data processing and dissemination [2]. Estrin et. al [3] worked on an hierarchical clustering mechanism dealing with asymmetric communication for power saving in sensor nodes. Jiang et.al presented a cluster based routing protocol (CBRP) [4]. According to this mechanism, all participating nodes of network are distributed in 2-hop cluster. Though this protocol is not much energy efficient for wireless sensor nodes however, it gives way to hierarchical clustering algorithms. Clustering for energy conservation is proven as efficient mechanism for wireless sensor networks [5,6]. When a sensor network is deployed, nodes establish clusters and nominate one node from each cluster as a cluster head. These cluster head nodes are responsible for receiving data from other nodes of cluster, do data aggregation/ fusion of received data and transmit it to base station. In this way, bandwidth consumption and life time of network is optimized [7]. In [8] authors give concept of inter cluster communication. They prove that regardless of transmitting fused data direct from cluster head to base station, if data is transmitted in multiple hopes i.e. from one cluster head to another and finally to base station, it would further enhance network life time.\\
Considering cluster based algorithms, today numerous protocols are developed, each having different attributes and enhancements mainly in cluster head selection algorithms. Though one thing is common, all protocols focus on energy conservation and data aggregation. M. Tahir \emph{et.al}[21] introduces link quality metric to divide a network into three logical portions resulting in lower routing overhead. Authors of [22] preserve energy in WSN's by differentiating idle and operational mode of a sensor node.\\
Authors of [9, 10] states that nodes having high initial energy will be selected as cluster heads (in case of heterogeneous sensor networks). While according [11, 12, 13] any node that lie within network can be elected as a cluster head. Stable Election Protocol (SEP) gives weighted probability to each node of becoming a cluster head [11]. In DEEC [12] existing energy in node is election criteria of a node to become a cluster head.\\
LEACH [1], TEEN [14], SEP [11], DEEC [12] and PEGASIS [15] are prominent routing techniques for wireless sensor networks. Main procedure of electing a cluster head was given by LEACH and that is further enhanced by SEP and DEEC. TEEN introduces the concept of thresholds that gives good results in network life time by showing reactive nature. These thresholds can be implemented in any routing protocol to enhance its performance with respect to utility or application. Considering LEACH, the algorithm is divided into three parts, i.e. advertising phase, Cluster Set up phase and Scheduling phase.\\
Based on LEACH, SEP and DEEC, numerous protocols are proposed. Q-LEACH [16] optimize network life time of homogenous wireless sensor network. [18] gives a detailed comparison analysis on different variants of LEACH as A-LEACH, S-LEACH and M-LEACH in terms of energy efficiency and applications. A very interesting comparison analysis between LEACH, Multi level Hierarchal LEACH and Multi hop LEACH is undertaken in [23]. Authors of [17] enhances SEP in terms of heterogeneity. They propose a model that gives three level heterogeneity. Whereas [19] gives a new protocol that works better than SEP in terms of network stability and life time having two level heterogeneity. T.N. Qureshi \emph{et.al} [20] modified DEEC protocol in terms of network stability, throughput as well as network life time.

\section{Motivation}
LEACH gives birth to many protocols. The procedures of this protocol are compact and well coped with homogeneous sensor environment. According to this protocol, for every round, new cluster head is elected and hence new cluster formation is required. This leads to unnecessary routing overhead resulting in excessive use of limited energy. If a cluster head has not utilized much of its energy during previous round, than there is probability that some low energy node may replace it as a cluster head in next cluster head election process. There is a need to limit change of cluster heads at every round considering residual energy of existing cluster head. Hence an efficient cluster head replacement algorithm is required to conserve energy.\\
In clustering protocols as LEACH, nodes use same amplification energy to transmit data regardless of distance between transmitter and receiver. To preserve energy, there should also be a transmission mechanism that specify required amplification energy for communicating with cluster head or base station. For example, transmitting a packet to cluster head with same amplification power level as required by a node located at farthest end of network to base station results in wastage of energy. One solution can be having global knowledge of network and than nodes decide how much they need to amplify signal. Locating and calculating distances with in full network topology needs lot of routing and so, this approach do not work for saving energy. To solve above mentioned problems, we propose two mechanisms. i.e. efficient cluster head replacement and dual transmitting power levels.\\

\section{Proposed Scheme}
Our work is based on LEACH protocol that can be extended to SEP and DEEC. Basically, we introduce two techniques to raise network life time and throughput. To understand our proposed scheme, we have to understand mechanism given by LEACH. This protocol changes the cluster head at every round and once a cluster head is formed, it will not get another chance for next $1/p$ rounds. For every round, cluster heads are replaced and whole cluster formation process is undertaken. \\
We, in this work, modify LEACH by introducing \emph{``efficient cluster head replacement scheme''}. It is a threshold in cluster head formation for very next round. If existing cluster has not spent much energy during its tenure and has more energy than required threshold, it will remain cluster head for the next round as well. This is how, energy wasted in routing packets for new cluster head and cluster formation can be saved. If cluster head has less energy than required threshold, it will be replaced according to LEACH algorithm.\\
Besides limiting energy utilization in cluster formation, we also introduce two different levels of power to amplify signals according to nature of transmission. Basically there can be three modes of transmission in a cluster based network.
\begin{enumerate}
  \item {Intra Cluster Transmission}
  \item {Inter Cluster Transmission}
  \item {Cluster Head To Base Station Transmission}
\end{enumerate}
Intra Cluster Transmission deals with all the communication within a cluster i.e. cluster members sense data and report sensed data to cluster head. The transmission/ reception between two cluster heads can be termed as inter cluster transmission while a cluster head transmitting its data straight to base station lies under the caption of cluster head to base station transmission.\\
Minimum amplification energy required for inter cluster or cluster head to BS communication and amplification energy required for intra cluster communication can not be same. In LEACH, amplification energy is set same for all kinds of transmissions. Using low energy level for intra cluster transmissions with respect to cluster head to BS transmission leads in saving much amount of energy. More over, multi power levels also reduce the packet drop ratio, collisions and/ or interference for other signals. In this context, we assume that a cluster at maximum may spread into an area of $10X10 m^2$ in a field of $100X100 m^2$. Energy that is enough to transmit at far ends of a field of $100 X 100 m^2$ must be lowered $10$ times for intra-cluster transmission. When a node act as a Cluster head, routing protocol informs it to use high power amplification and in next round, when that node becomes a cluster member, routing protocol switches it to low level power amplification. Finally, soft and hard threshold schemes are also implemented in MODLEACH that gives better results.
\begin{figure*}
\begin{center}
\includegraphics[width=14cm,height=16cm]{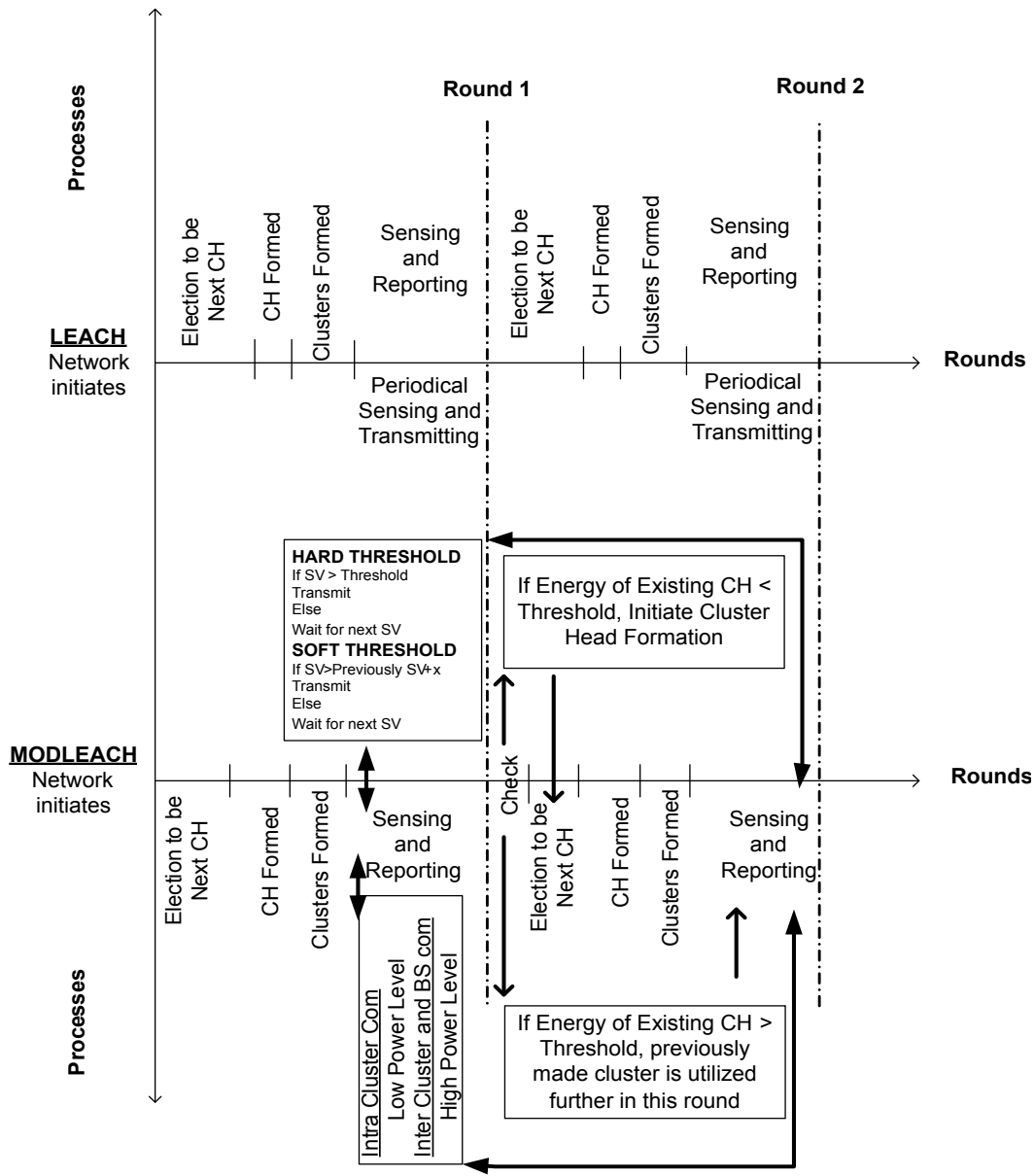}
\caption{Protocol Functioning}
\end{center}
\end{figure*}
\section{Experiments, Results and Discussion}

\begin{table}
  \centering
 \caption{Network Parameters}
\begin{tabular}{p{6cm}| p{2.5cm}}
\hline{}
 \textbf{Network Parameters }              & \textbf{Value}	    \\
\hline{}
Network Size  	& $ 100 X 100 m^2 $  \\
Initial Energy of Sensor Nodes &	0.5 J\\
Packet Size	& 4000 bits \\
Transceiver idle state energy consumption	& 50 nJ/bit \\
Data Aggregation/ Fusion Energy consumption&	5 nJ/bit/report\\
Amplification Energy (Cluster to BS)$d\geq d_o$	& $Efs=10pJ/bit/m^2$ \\
Amplification Energy (Cluster to BS) $d \leq d_o$ &	$Emp=0.0013pJ/bit/m^2$\\
Amplification Energy (Intra Cluster Comm.) $d\geq d_1$	& $Efs/10= Efs_1$\\
Amplification Energy (Intra Cluster Comm.)$d \leq d_1$ &	$Emp/10= Emp_1$\\


 \end{tabular}

\end{table}

Simulations are conducted using $ MATLAB (R2009a)$ and to get precise plots, confidence interval is taken. Simulations show that MODLEACH performs better considering metrics of throughput, network life time, and optimized cluster head formation of network. MODLEACH is further improved by using the concept of soft and hard threshold as introduced by TEEN. MODLEACHHT further improve efficiency however, MODLEACHST performs best amongst all.

\subsubsection{Network Life Time}

Considering network life time of LEACH, MODLEACH, MODLEACHHT and MODLEACHST, LEACH has lowest performance with respect to network life time. MODLEACH has greater stable period due to its efficient cluster head replacement scheme and dual transmitting power level for inter and intra cluster communication. Simulated results depicted in figure $2$ and figure $3$ represent network life time by showing number of alive and dead nodes respectively.

\begin{figure}
\begin{center}
\includegraphics[scale=0.28]{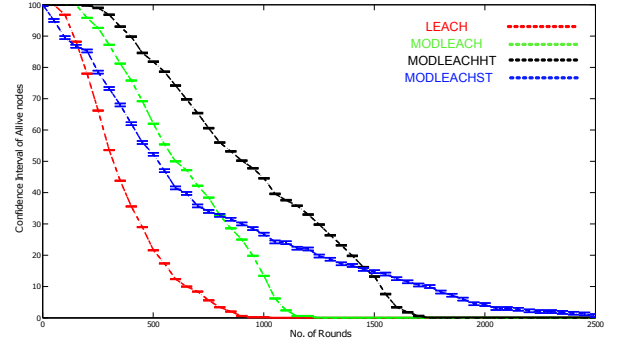}
\caption{Comparison, Alive Nodes of Network}
\end{center}
\end{figure}

MODLEACHST gives maximum network life time amongst all protocols. This due to limiting number of transmissions (concept of soft threshold) along with efficient cluster head replacement mechanism that preserve energy globally and multi power level for inter and intra cluster communication. In MODLEACHST, number of transmissions are confirmed only when a pre-described change in sensed data is achieved. This limits number of transmissions to preserve residual energy of a sensor node (numbers of transmissions are inversely proportional to energy of sensor node). Considering MODLEACHHT, it performs 2nd best with respect to network life time as it implements the concept of hard threshold. When a sensed value is greater than a threshold level, data packet will only be transmitted then. This also deals with minimizing number of transmissions to save energy of a node. However, it transmits data always when that threshold is broken; no matter if sensor is sensing the same value from last many rounds.

\begin{figure}
\begin{center}
\includegraphics[scale=0.28]{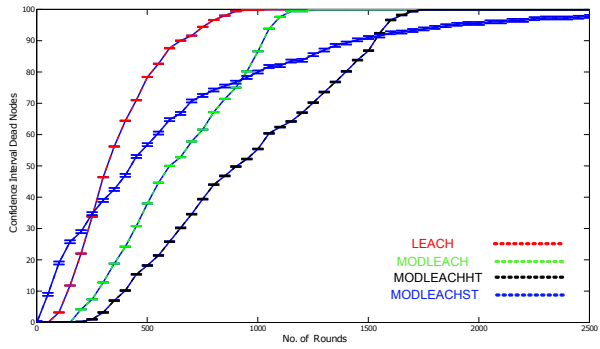}
\caption{Comparison, Dead Nodes of Network/ Network Life Time}
\end{center}
\end{figure}

\subsubsection{Throughput}

\begin{figure}
\begin{center}
\includegraphics[scale=0.28]{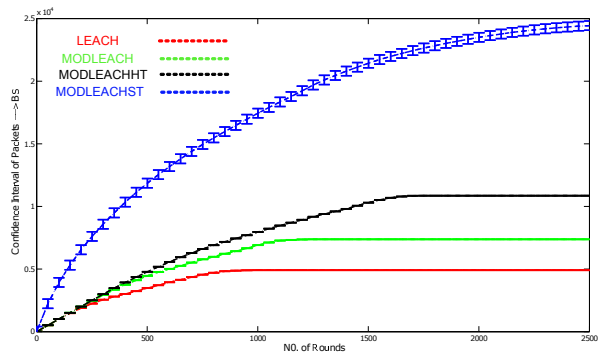}
\caption{Comparison, Packets Transmitted to Base Station}
\end{center}
\end{figure}

Besides network life time, another metric to judge efficiency of a routing protocol is its throughput. A base station receiving more data packets confirms the efficiency of routing protocol. Throughput depends on network life time in a sense but not always. Considering the simulated results as shown in figure $4$, we deduce that, maximum throughput is achieved by MODLEACHST. Better network life time and efficient cluster head replacement mechanism are two major reasons of increased throughput in MODLEACHST. Comparing MODLEACH and LEACH, MODLEACH gives better throughput due to same reasons i.e. increased network life time and better cluster head replacement scheme. One more prominent reason is dual transmitting power levels with in network. Using different amplification energies for transmissions reduce packet drop ratio resulting in higher through put. Another reason that can distinguish throughput gain in all studied techniques is the mode of operation. LEACH and MODLEACH both are proactive (periodical transmissions) in nature while MODLEACHHT and MODLEACHST are reactive (event driven). This also depicts that proactive (periodical transmission) routing protocols have lower throughput than reactive (event driven) routing protocols.

\begin{figure}
\begin{center}
\includegraphics[scale=0.28]{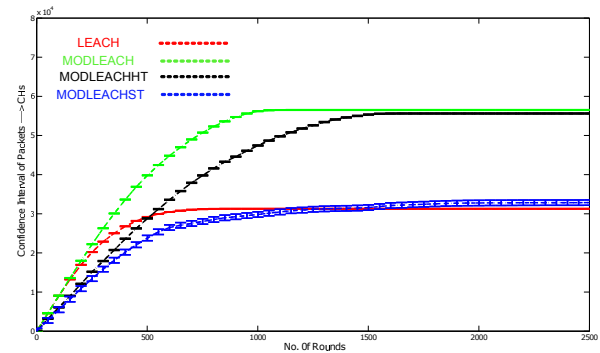}
\caption{Comparison, Packets Transmitted to Cluster Heads}
\end{center}
\end{figure}

\subsubsection{Cluster Head Formation and Scope}

\begin{figure}
\begin{center}
\includegraphics[scale=0.28]{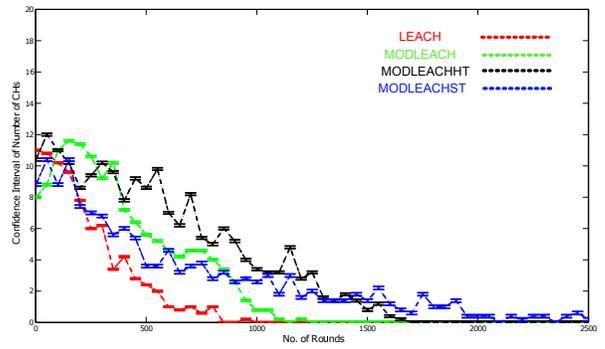}
\caption{Comparison, Number of  Cluster Heads}
\end{center}
\end{figure}

Figure $6$ and Figure $7$ show number of cluster heads chosen at each round. All the techniques basically used same algorithm hence no major difference is there in cluster head formation and calculation manner however, MODLEACH differs from LEACH in a sense that initially its number of cluster heads remain stable and then cluster head formation behavior goes similar to that of LEACH. Initially as stated above, MODLEACH executes same cluster heads for next round/s if they have energy greater than defined threshold. This is the reason of stable number of cluster heads initially.

\begin{figure}
\begin{center}
\includegraphics[scale=0.35]{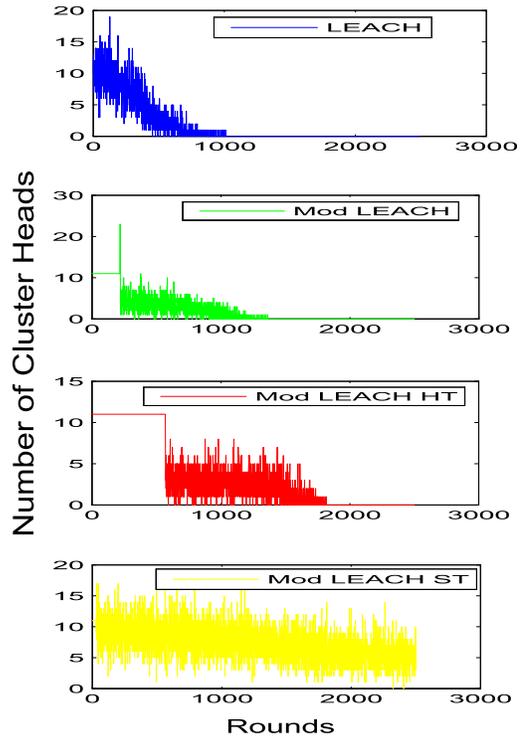}
\caption{Cluster Head Formation per Round}
\end{center}
\end{figure}

\section{Conclusion}
In this work, we give a brief discussion on emergence of cluster based routing in wireless sensor networks. We also propose MODLEACH, a new variant of LEACH that can further be utilized in other clustering routing protocols for better efficiency. MODLEACH tends to minimize network energy consumption by efficient cluster head replacement after very first round and dual transmitting power levels for intra cluster and cluster head to base station communication. In MODLEACH, a cluster head will only be replaced when its energy falls below certain threshold minimizing routing load of protocol. Hence, cluster head replacement procedure involves residual energy of cluster head at the start of each round. Further, soft and hard thresholds are implemented on MODLEACH to give a comparison on performances of these protocols considering throughput and energy utilization.
\\In future, we will carry our work to calculate routing load of MODLEACH, MODLEACHST and MODLEACHHT analytically and apply efficient cluster head replacement mechanism along with dual transmission power levels in other clustering routing protocols of wireless sensor networks to study their impact in a broader sense.

\end{document}